\documentclass[
 reprint,
 amsmath,amssymb,
 aps,
 prl
]{revtex4-2}

\usepackage{amsmath}
\usepackage{graphicx, color}
\usepackage{dcolumn}
\usepackage{bm}
\usepackage[utf8]{inputenc}

\newcommand{\nn}{\nonumber}
\newcommand{\beq}{\begin{equation}}
\newcommand{\eeq}{\end{equation}}
\newcommand{\beqa}{\begin{eqnarray}}
\newcommand{\eeqa}{\end{eqnarray}}

\newcommand{\Bbar}{\,\overline{\!B}{}}
\newcommand{\Dbar}{\,\overline{\!D}{}}
\newcommand{\Kbar}{\,\overline{\!K}{}}
\def\B0bar{\Bbar{}^0}
\def\D0bar{\Dbar{}^0}
\def\K0bar{\Kbar{}^0}

\def\lmd{\lambda}
\def\eps{\epsilon}
\def\Heff{\mathcal{H}_{\rm eff}}
\def\O{\mathcal{O}}
\def\cbar{\overline{c}}

\def\ellbar{\overline{\ell}}

\graphicspath{{ps}}

\def\gf#1{\langle g_{#1}\rangle}

\begin{document}

\preprint{}

\title{Unbinned Angular Analysis of $B\to D^*\ell \nu_\ell$ and the Right-handed Current}

\author{Z.R.~Huang}
 \email{zhuang@lal.in2p3.fr}
\author{E.~Kou}
 \email{kou@lal.in2p3.fr}
\affiliation{Université Paris-Saclay, CNRS/IN2P3, IJCLab, 91405 Orsay, France}

\author{C.D.~L\"u}
\email{lucd@ihep.ac.cn}
\author{R.Y.~Tang}
\email{tangruying@ihep.ac.cn}
\affiliation{Institute of High Energy Physics, Chinese Academy of Sciences, Beijing 100049, China\\}
\affiliation{School of Physics, University of Chinese Academy of Sciences, Beijing 100049, China}

\date{\today}

\begin{abstract}
In this article, we perform a sensitivity study of an unbinned angular analysis of the $B\to D^*\ell \nu_\ell$ decay, including the contributions from the right-handed current. We show that the angular observable can constrain very strongly the right-handed current without the intervention of the yet unsolved $V_{cb}$ puzzle.
\end{abstract}

\maketitle

\section{\label{sec:1}Introduction}
The $B\to  D^* \ell\nu_\ell$ ($\ell= e$ or $\mu$) has been receiving a great deal of attention in recent years.
Two main reasons are the so-called $V_{cb}$ puzzle and the $R(D^{(*)})$ anomalies.
The former is the problem that there is a tension between the values of $|V_{cb}|$ determined by using
the experimental measurements of the inclusive and the exclusive $b\to c \ell \nu_\ell$ decays.
The $|V_{cb}|$ determination from the exclusive $B\to  D^{(*)} \ell\nu_\ell$  decay depends on the $B\to D^{(*)}$ form factors.
Simultaneous measurements~\cite{Waheed:2018djm,Abdesselam:2017kjf,BaBar:2019vpl} of the $|V_{cb}|$ and the form factors using the angular distribution~\cite{Bhattacharya:2020lfm,Duraisamy:2013pia,Iguro:2020keo,Ivanov:2017mrj,Ivanov:2015tru,Becirevic:2016hea} of the $B\to  D^{*} \ell\nu_\ell$ have been attempted, which were followed by intensive theoretical interpretations~\cite{Bernlochner:2017xyx,Bernlochner:2019ldg,Bigi:2017njr,Gambino:2019sif,Grinstein:2017nlq,Iguro:2020cpg,Jaiswal:2020wer,Jung:2018lfu,Bordone:2019guc,Bordone:2019vic,Bobeth:2021lya,Ricciardi:2019zph,Colangelo:2018cnj,Bhattacharya:2019olg}. The most exciting progress we foresee in this study is that new average of lattice QCD results will become possible~\cite{Bazavov:2021bax}. The latter $R(D^{(*)})$ anomalies are the discrepancies between the Standard Model (SM) predictions and the experimental results on the ratios of $\mathcal B(B\to  D^{(*)} \tau\nu_\tau)$ and $\mathcal B(B\to  D^{(*)} \ell\nu_\ell)$. Various investigations assuming that this is an appearance of the physics beyond the SM are ongoing~\cite{Fajfer:2012jt,Sakaki:2013bfa,Murgui:2019czp,Becirevic:2012jf,Cheung:2020sbq,Huang:2018nnq,Li:2016vvp,Crivellin:2012ye,Altmannshofer:2017poe,Asadi:2018sym,Jaiswal:2017rve}. In order to confirm that these phenomena are indeed the new physics discoveries, detailed studies need to be carried out both theoretically and experimentally.

In this article, motivated by these phenomenological problems, we examine the usefulness of {\it the unbinned angular distribution measurements} to scrutinise the $B\to  D^* \ell\nu_\ell$ decay. The existing experimental analysis mentioned above utilised four one dimensional binned distributions: they are the projections of one of the three angles ($\theta_l, \theta_V, \chi$) and one momentum ($w$). On the other hand, once a larger amount of data becomes available at the Belle II experiment~\cite{Kou:2018nap}, the unbinned analysis with simultaneous fit of three angular distributions will become possible.  The method is similar to the one which was applied for the $B\to K^* \mu^+\mu^-$ decay where another anomaly is found~\cite{Aaij:2015oid}. We expect that the angular distribution would be most useful to distinguish the new physics contributions which carry opposite chirality to the SM, i.e. the right-handed vector contribution~\cite{Crivellin:2009sd,Crivellin:2014zpa,Alioli:2017ces}, which can be induced in some NP scenarios, e.g. from the $W_L-W_R$ mixing in the left-right symmetric model~\cite{Kou:2013gna}. Furthermore, in \cite{Cata:2015lta,Cirigliano:2009wk}, it is pointed out that the right-handed vector current contribution is lepton flavour universal at tree level in the context of the linear electroweak symmetry breaking. Thus, the NP effects in $B\to  D^* \ell\nu_\ell$ may have a strong implication for $B\to  D^* \tau\nu_\tau$ and $B_c\to \tau \nu_\tau$ processes.

In this article, we will investigate the impact of the unbinned angular distribution measurements to the investigation of new physics solely from the right-handed vector current involving a light charged lepton. We will utilize the pseudodata generated using hadronic form factors from the Belle analysis~\cite{Waheed:2018djm} for the light modes and discuss the role of the lattice QCD results which will become available soon.
\section{\label{sec:2}The unbinned angular analysis}
The weak Hamiltonian for $B \to D^*\ell\nu_\ell$ decay including the left-handed (SM) operator and the right-handed operator (assuming no right-handed neutrino) is
\begin{small}
\begin{equation}
   \Heff ={4G_F \over \sqrt2} V_{cb}\left[ C_{V_L}\O_{V_L} + C_{V_R}\O_{V_R} \right] + \text{h.c.} \,,
      \label{eq:Ham}
\end{equation}
\end{small}
where $G_F$ and $V_{cb}$ are respectively the Fermi Constant and the Cabibbo-Kobayashi-Maskawa (CKM) matrix element, and $C_{V_L}$ and $C_{V_R}$ are the Wilson coefficients of the left-handed and the right-handed vector operators with $\O_{V_L}$ and $\O_{V_R}$ defined as
\begin{small}
\begin{equation}
 \O_{V_L} = (\cbar_L \gamma^\mu b_L)(\ellbar_L \gamma_\mu \nu_{L}) \,,  \,\,\,
     \O_{V_R} = (\cbar_R \gamma^\mu b_R)(\ellbar_L \gamma_\mu \nu_{L}) \,.
   \label{eq:operators}
\end{equation}
\end{small}
In the SM, we have $C_{V_L}=1$ and $C_{V_R}=0$, while in some NP scenarios such as the left-right symmetric model~\cite{Kou:2013gna}, $C_{V_R}$ can be non-zero.

Having the above weak Hamiltonian, let us then build our probability density function (PDF) in terms of 11 independent angular observables $J_i$, defined as functions of Wilson coefficients and helicity amplitudes in Eq.~\eqref{eq:rateJ} and \eqref{eq:Jfunction} in Appendix. First, we integrate out all the angles to obtain the normalisation~\cite{Bernlochner:2014ova}:
\begin{small}
\begin{eqnarray}
\frac{d\Gamma}{dw} &=& \frac{6m_Bm_{D^*}^2}{8(4\pi)^4}G_F^2 \,\eta_{\rm EW}^2 \left|V_{cb}\right|^2 \, \times \mathcal{B}(D^{*} \to D \pi)  \nn \\
 &\times &\frac{8\pi}{9} \Big\{6J_{1s}^{\prime}+3J_{1c}^{\prime}-2J_{2s}^{\prime}-J_{2c}^{\prime}\Big\}\,,
\end{eqnarray}
\end{small}
where $J^{\prime}_i \equiv J_i\sqrt{w^2-1}(1-2wr+r^2)$. In the following, to take into account the $w$ dependence, we separate $w$ in 10 bins and prepare the PDF for each bin. We express the decay rate for each bin as
\begin{small}
\begin{eqnarray}
&&\langle \Gamma\rangle_{w-{\rm bin}} =  \frac{6m_Bm_{D^*}^2}{8(4\pi)^4}G_F^2 \,\eta_{\rm EW}^2\, \left|V_{cb}\right|^2 \, \times \mathcal{B}(D^{*} \to D \pi)  \nn \\
&\times & \frac{8\pi}{9} \Big\{6\langle J_{1s}^{\prime}\rangle_{w-{\rm bin}}+3\langle J_{1c}^{\prime}\rangle_{w-{\rm bin}}-2\langle J_{2s}^{\prime}\rangle_{w-{\rm bin}}-\langle J_{2c}^{\prime}\rangle_{w-{\rm bin}} \Big\}\,.\nn\\
\label{eq:Gbin}
\end{eqnarray}
\end{small}

Hereafter, the index $w$-bin is implicit. Now, the PDF is written by new {\it normalised} angular coefficients $g_i$ as:
\begin{small}
\begin{eqnarray}
&&\hat{f}_{\vec{\gf{}}}(\cos\theta_V, \cos\theta_\ell, \chi)=\frac{9}{8\pi} \nn \\
&\times& \Big\{ \frac{1}{6}(1-3 \gf{1c} +2\gf{2s}+\gf{2c}) \sin^2\theta_V+\gf{1c} \cos^2\theta_V \nn \\
&\quad& +(\gf{2s} \sin^2\theta_V+\gf{2c}\cos^2\theta_V )\cos 2\theta_\ell \nn \\
&\quad &+\gf{3} \sin^2\theta_V\sin^2\theta_\ell\cos 2\chi \nn \\
&\quad & +\gf{4}\sin 2\theta_V\sin 2\theta_\ell \cos\chi
+\gf{5} \sin 2\theta_V\sin\theta_\ell\cos\chi \nn\\
&\quad &+(\gf{6s} \sin^2\theta_V+\gf{6c}\cos^2\theta_V)\cos\theta_\ell \nn \\
&\quad &+\gf{7} \sin 2\theta_V\sin\theta_\ell \sin\chi+\gf{8}\sin 2\theta_V \sin 2\theta_\ell\sin\chi \nn \\
&\quad &+\gf{9} \sin^2\theta_V\sin^2\theta_\ell \sin2\chi\Big\}\,,
\end{eqnarray}
\end{small}
where
\begin{small}
\begin{equation}
\gf{i} \equiv \frac{\langle J_i^{\prime}\rangle }{6\langle J_{1s}^{\prime}\rangle+3\langle J_{1c}^{\prime}\rangle-2\langle J_{2s}^{\prime}\rangle-\langle J_{2c}^{\prime}\rangle}\,.
\end{equation}
\end{small}
Notice that $\gf{6s}$ is equivalent to the Forward-Backward Asymmetry (FBA) up to a constant:
\begin{small}
\begin{eqnarray}
{\langle \mathcal{A}_{\theta_\ell} \rangle} &\equiv& \frac{\int_0^{1} {d\Gamma\over d\cos\theta_\ell}  {\text{d}}\cos\theta_\ell -\int_{-1}^{0} {d\Gamma\over d\cos\theta_\ell}  {\text{d}}\cos\theta_\ell}{\int_0^{1} {d\Gamma\over d\cos\theta_\ell} {\text{d}}\cos\theta_\ell +\int_{-1}^{0} {d\Gamma \over d\cos\theta_\ell} {\text{d}}\cos\theta_\ell} \nn \\
&=& 3\gf{6s}\,.
\end{eqnarray}
\end{small}

Now having the PDF, the experimental determination of the $\gf{i}$ can be pursued by the maximum likelihood method:
\begin{small}
\begin{equation}
\mathcal{L}(\vec{\gf{}})=\sum_{i=1}^{N} \ln \hat{f}_{\vec{\gf{}}} (e_i)\,, \label{eq:55}
\end{equation}
\end{small}
where $e_i$ indicates the experimental events and $N$ is the number of events.

The error matrix for $\gf{i}$ can be obtained via the covariance matrix, $V_{ij}$, which is the $11\times 11$ matrix for each $w$-bin (so we need 10 of these matrices if we have 10 bins). In this work, based on the truth values of $\gf{i}$ obtained using the measured form factors in \cite{Waheed:2018djm}, we use the toy Monte-carlo method to generate the covariance matrices.

Using the pseudodata of $\gf{i}$, the Wilson coefficient $C_{V_R}$ and the parameters in hadronic form factors can be fitted. Following the Belle analysis~\cite{Waheed:2018djm}, we use two sets of parametrisation for $B\to D^*$ form factors, i.e. the CLN parametrisation~\cite{Caprini:1997mu} based on heavy quark expansion (HQE) and the BGL parametrisation~\cite{Boyd:1997kz} based on analyticity, despite the fact that there is updated HQE parametrisation \cite{Bernlochner:2017jka,Bordone:2019vic} which is more flexible by including higher-order terms in $1/{m_{b,c}}$ expansion and z expansion. The theoretical parameters $\vec{v}$, which is $\vec{v}=(h_{A_1}(1), \rho^2_{D^*}, R_1(1), R_2(1), V_{cb})$ for the CLN parametrisation and  $\vec{v}=(a^g_{0,1,\cdots}, a^f_{0,1,\cdots}, a^{\mathcal F_1}_{1,2,\cdots}, V_{cb})$ for the BGL parametrisation~\cite{Boyd:1997kz} are fitted by minimising the following $\chi^2$:
\begin{small}
\begin{equation}
\chi^2(\vec{v})=\chi^2_{\rm angle}(\vec{v})+\chi^2_{w-{\rm bin}}(\vec{v})+\chi^2_{\rm lattice}(\vec{v})\,,
\label{eq:chi2sum}
\end{equation}
\end{small}
where $\chi^2_{\rm angle}(\vec{v})$ takes into account the angular distribution and $\chi^2_{w-{\rm bin}}(\vec{v})$ does the $w$ dependence. The $\chi^2_{\rm lattice}$ is the constraint from the lattice QCD computation, which we explain more in detail below.

The first term can be given as
\begin{small}
\begin{eqnarray}
&&\chi^2_{\rm angle}(\vec{v}) =\sum_{w-{\rm bin=1}}^{10}\Big[\sum_{ij}N_{\rm event}   \\
&&\quad\quad \hat{V}_{ij}^{-1}(\gf{i}^{\rm exp}-\langle g_i^{\rm th}(\vec{v}) \rangle )(\gf{j}^{\rm exp}-\langle g_j^{\rm th}(\vec{v}) \rangle)\Big]_{w-{\rm bin}}\,, \nn
\end{eqnarray}
\end{small}
where $\hat{V}$ and $\gf{i}^{\rm exp}$ are the covariance matrix and the $w$-bin integrated $g_i$ functions. The second term is given as
\begin{small}
\begin{equation}
\chi^2_{w-{\rm bin}}(\vec{v}) = \sum_{w-{\rm bin=1}}^{10} \frac{([N]_{w-{\rm bin}}-\alpha \langle \Gamma\rangle_{w-{\rm bin}} )^2}{[N]_{w-{\rm bin}}}\,,
\end{equation}
\end{small}
where  $\langle \Gamma\rangle_{w-{\rm bin}}$ is in Eq.~(\ref{eq:Gbin}).  In reality, there might be an experimental correlation between different $w$-bin, which must be taken into account.

The factor $\alpha$ is a constant, which relates the number of events and the decay rate:
\begin{small}
\begin{equation}
\alpha\equiv \frac{4N_{B\overline{B}}}{1+f_{+0}} \tau_{B^0}\times \epsilon{\mathcal{B}}(D^0\to K^-\pi^+)\,,
\end{equation}
\end{small}
where $N_{B\overline{B}}=(772\pm 11) \times 10^6$ is the number of $B\overline{B}$ pairs produced from $\Upsilon(4S)$, which corresponds to 711 fb$^{-1}$ of data at Belle~\cite{Belle:2000cnh}, $f_{+0}$ is the constant defined as $f_{+0}={\Gamma(\Upsilon(4S)\to B^+B^-)\over \Gamma(\Upsilon(4S)\to B^0{\overline B^0})}$, $\tau_{B^0}$ is the lifetime of $B^0$, and $\epsilon$ is the experimental efficiency (the values are from PDG~\cite{ParticleDataGroup:2020ssz}).

In the following sections, we will investigate the sensitivity of the unbinned angular analysis proposed above to  the right-handed current, i.e. $C_{V_R}$. We use the best-fit values of the CLN and BGL parameters in the Belle '18 paper~\cite{Waheed:2018djm}, in order to generate the {\it pseudo}-experimental data. The total number of events is also adjusted to $\sim 95$k as in~\cite{Waheed:2018djm}, corresponding to roughly the universal efficiency of $\epsilon \sim 4.8\times 10^{-2}$. Thus, the $\alpha$ parameter is computed as $6.616(6.613)\times10^{18}$ for CLN(BGL) parametrisation.

\begin{figure}[htbp]
\def\figwid{2.8}
\def\figscale{0.20}
\includegraphics[width=\figwid cm]{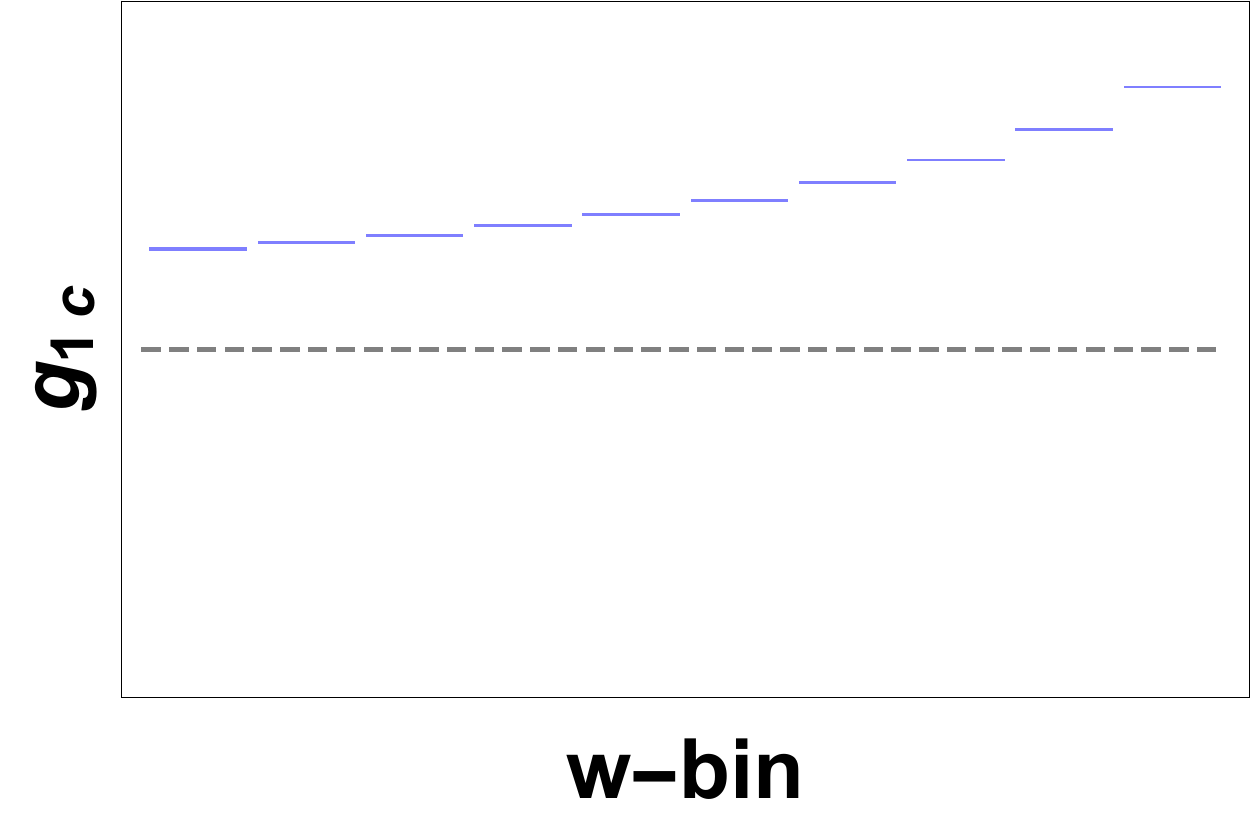}
\includegraphics[width=\figwid cm]{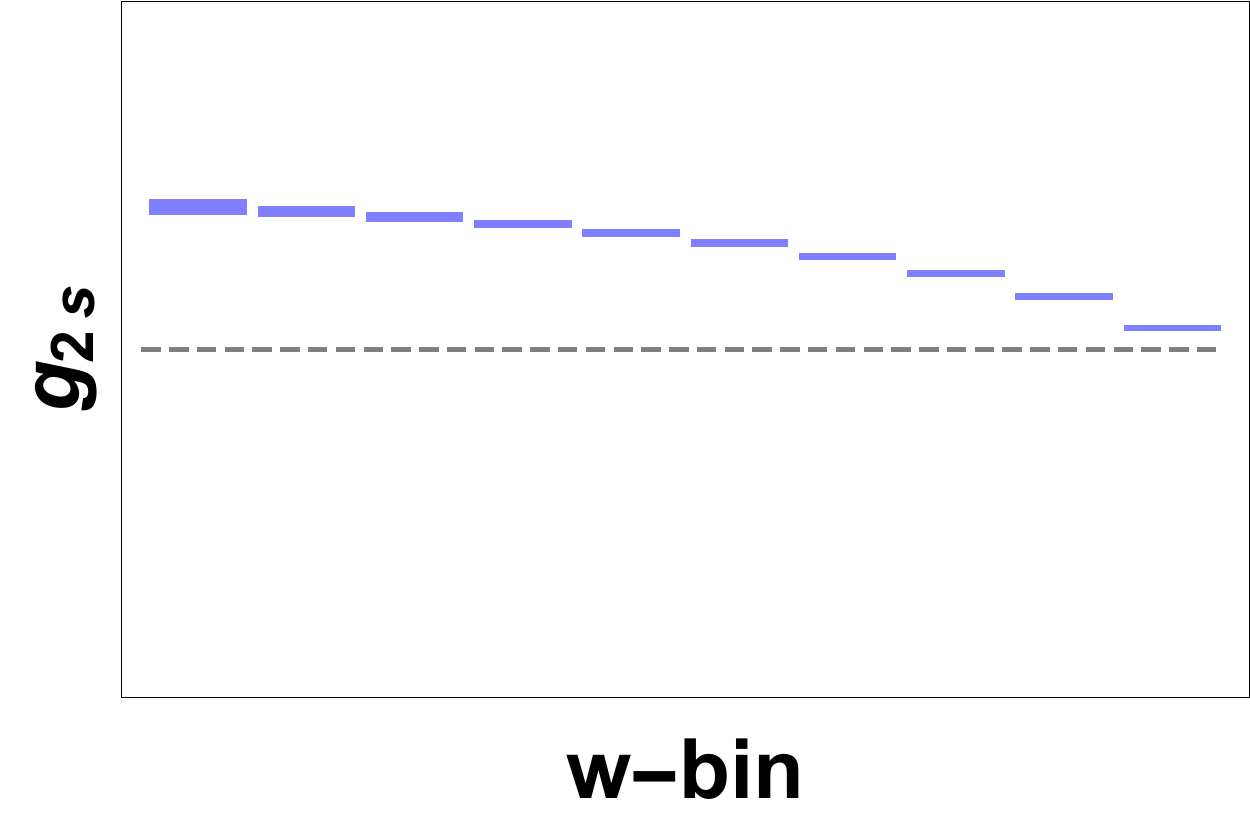}
\includegraphics[width=\figwid cm]{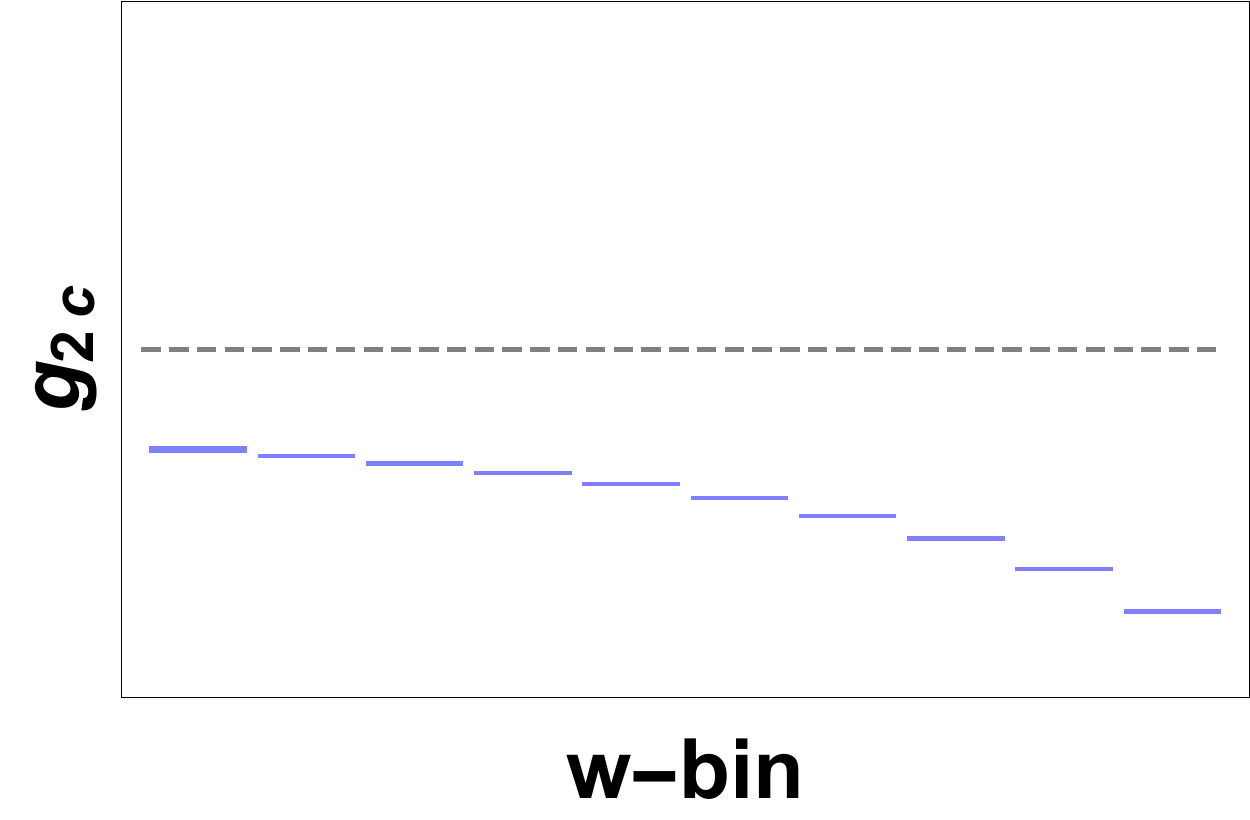}
\includegraphics[width=\figwid cm]{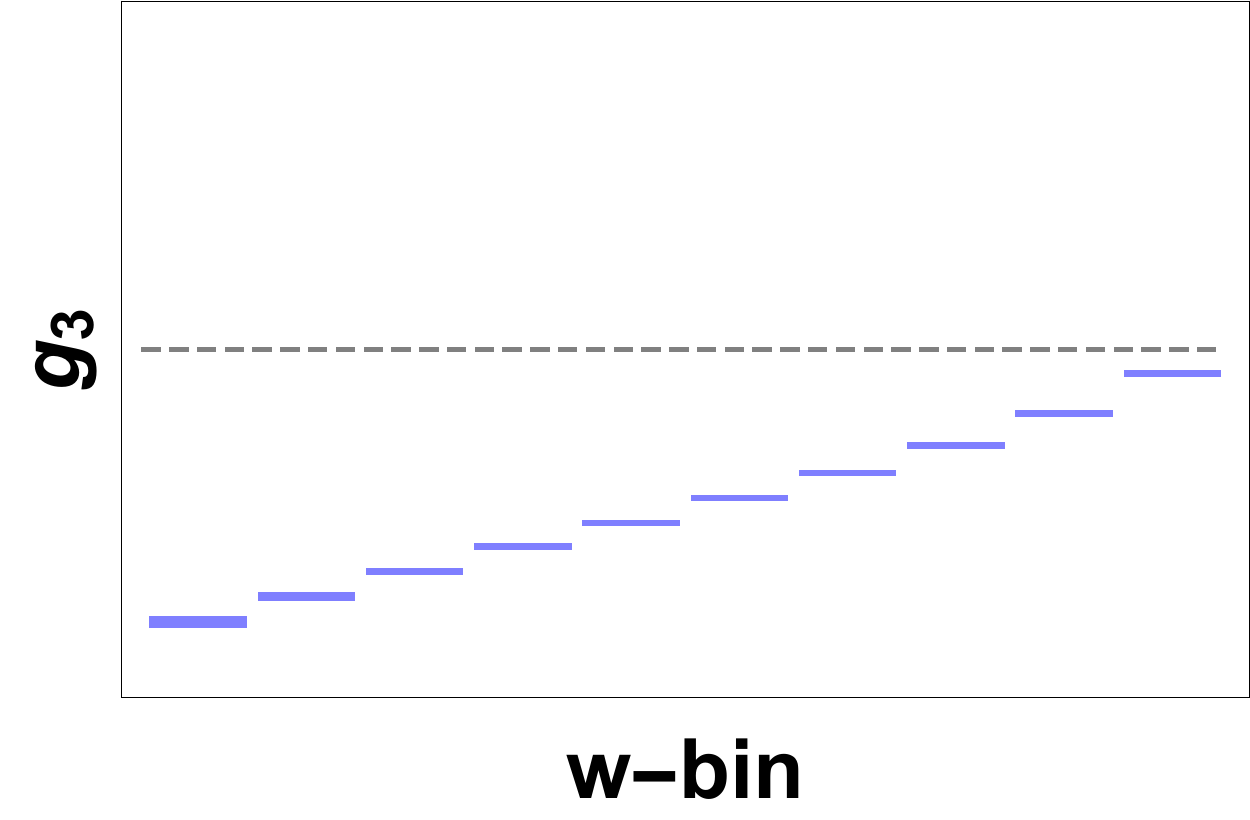}
\includegraphics[width=\figwid cm]{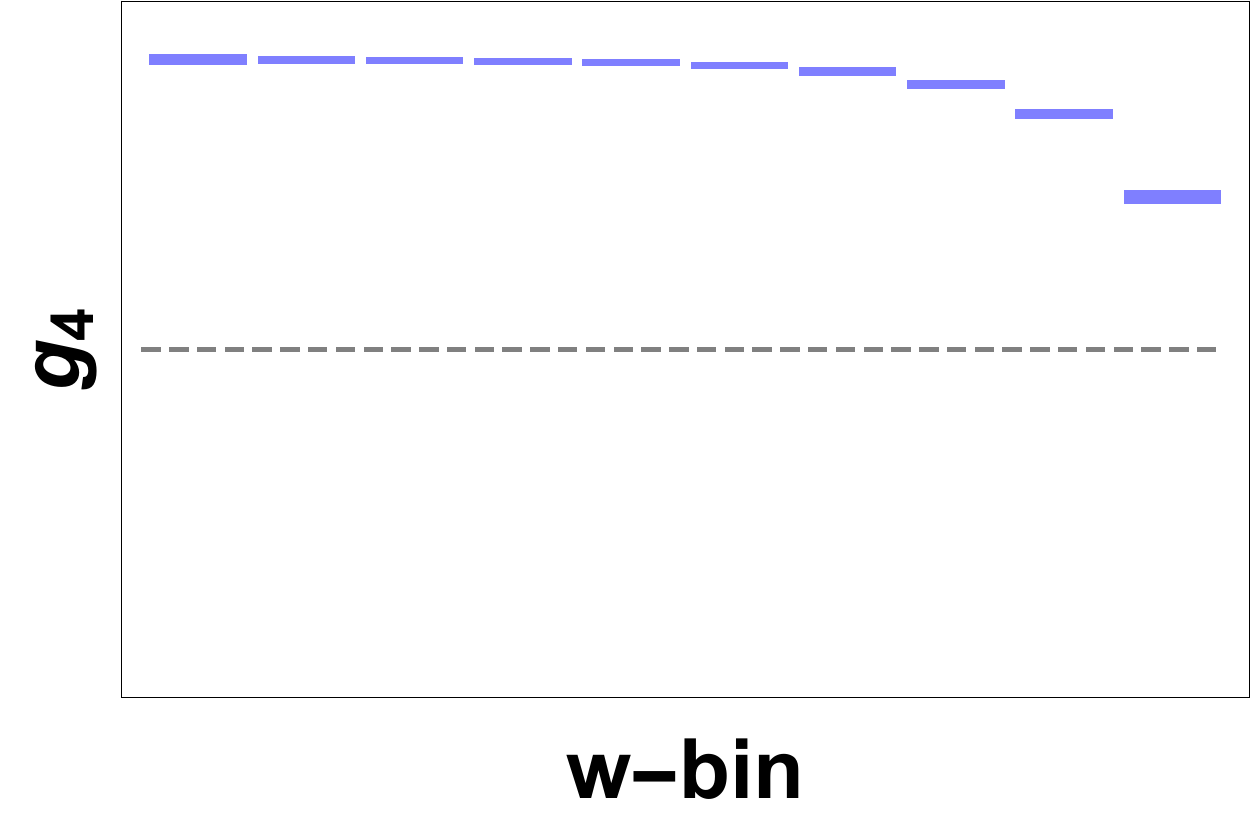}
\includegraphics[width=\figwid cm]{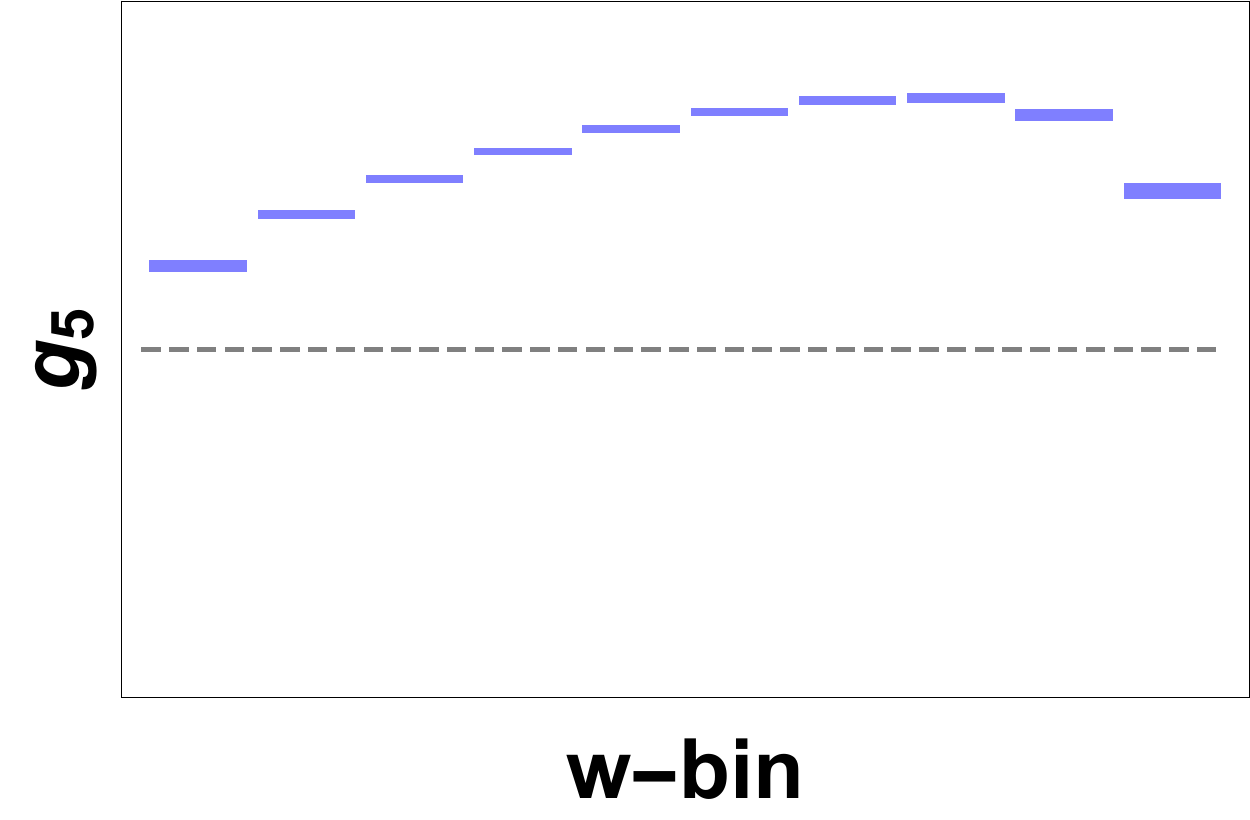}
\includegraphics[width=\figwid cm]{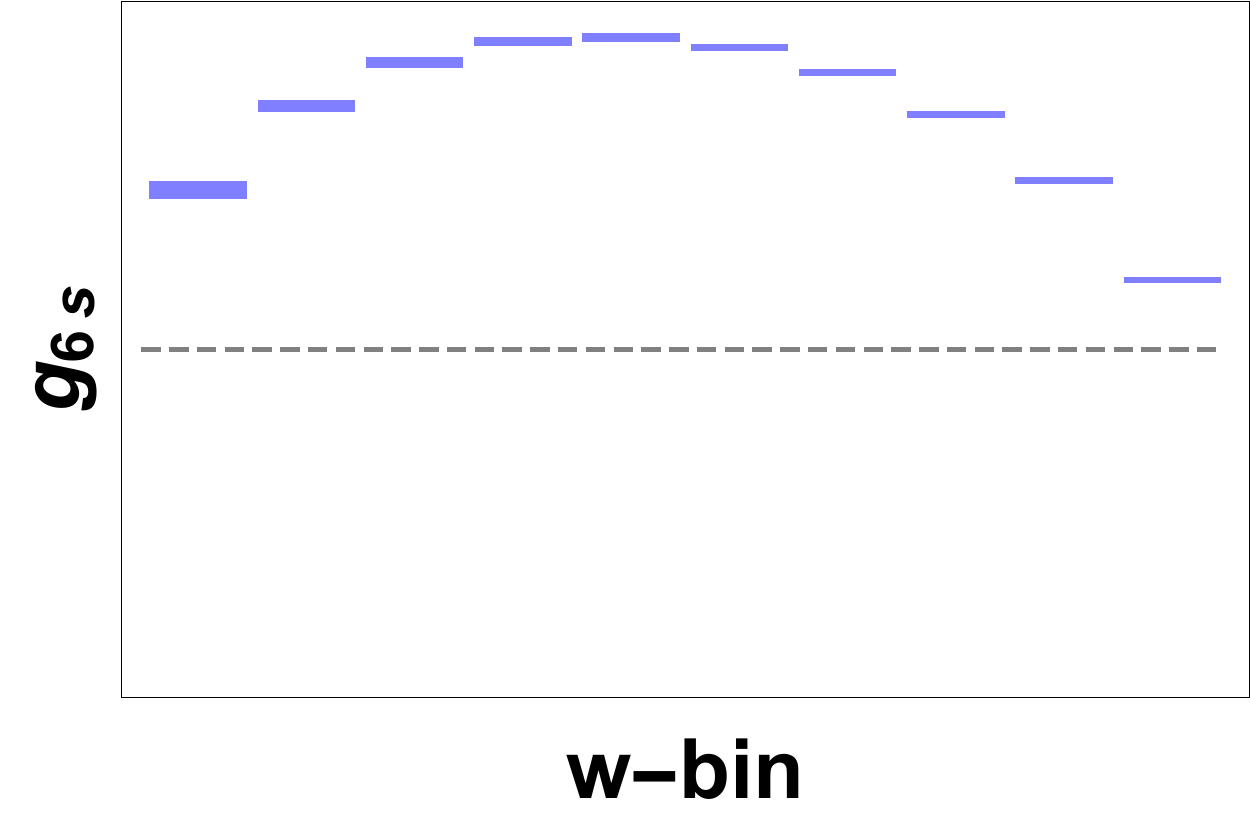}
\includegraphics[width=\figwid cm]{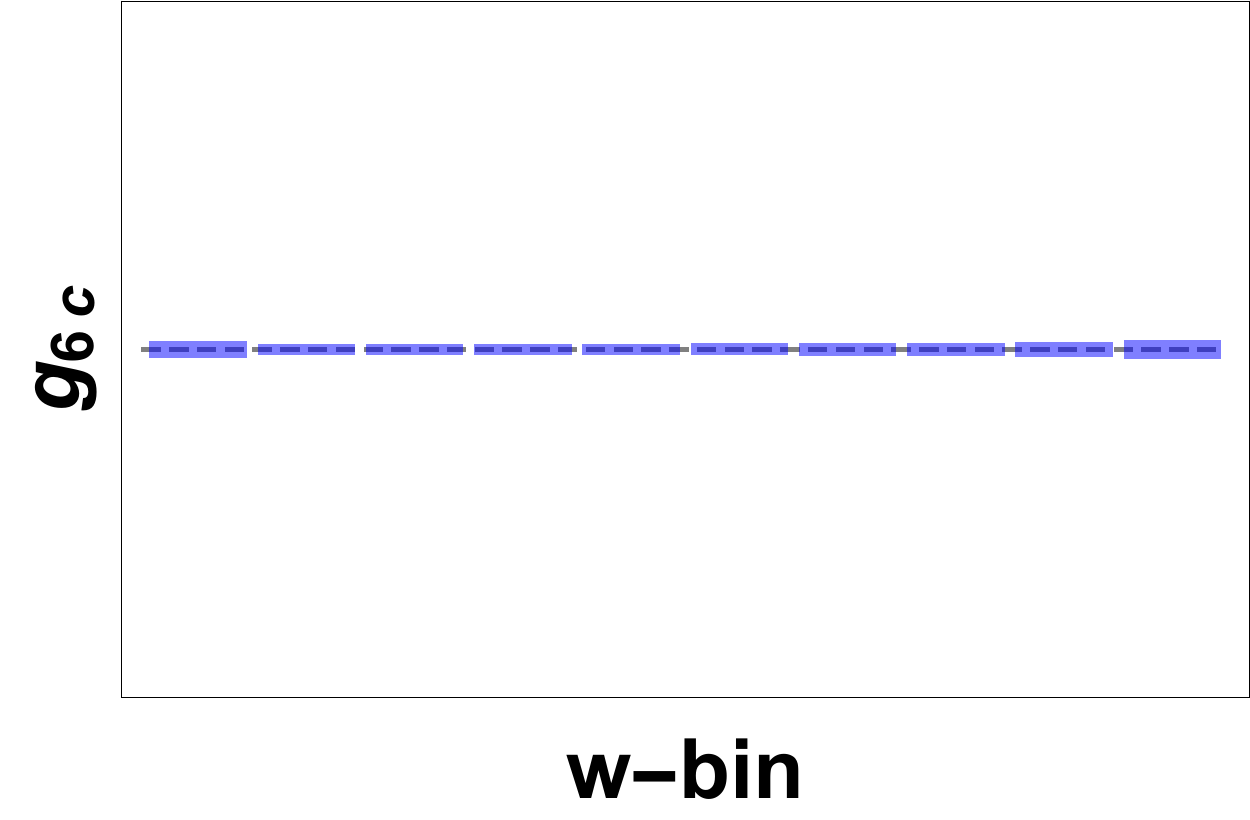}
\includegraphics[width=\figwid cm]{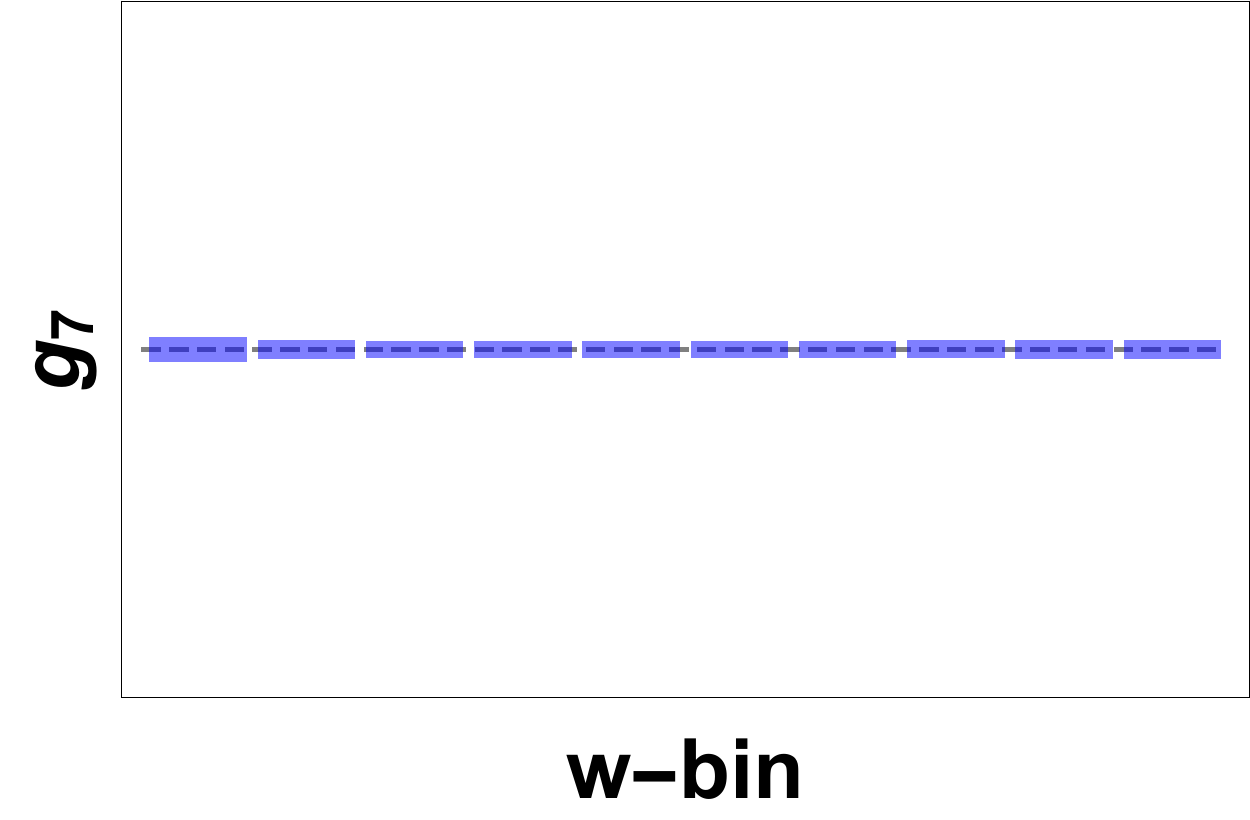}
\includegraphics[width=\figwid cm]{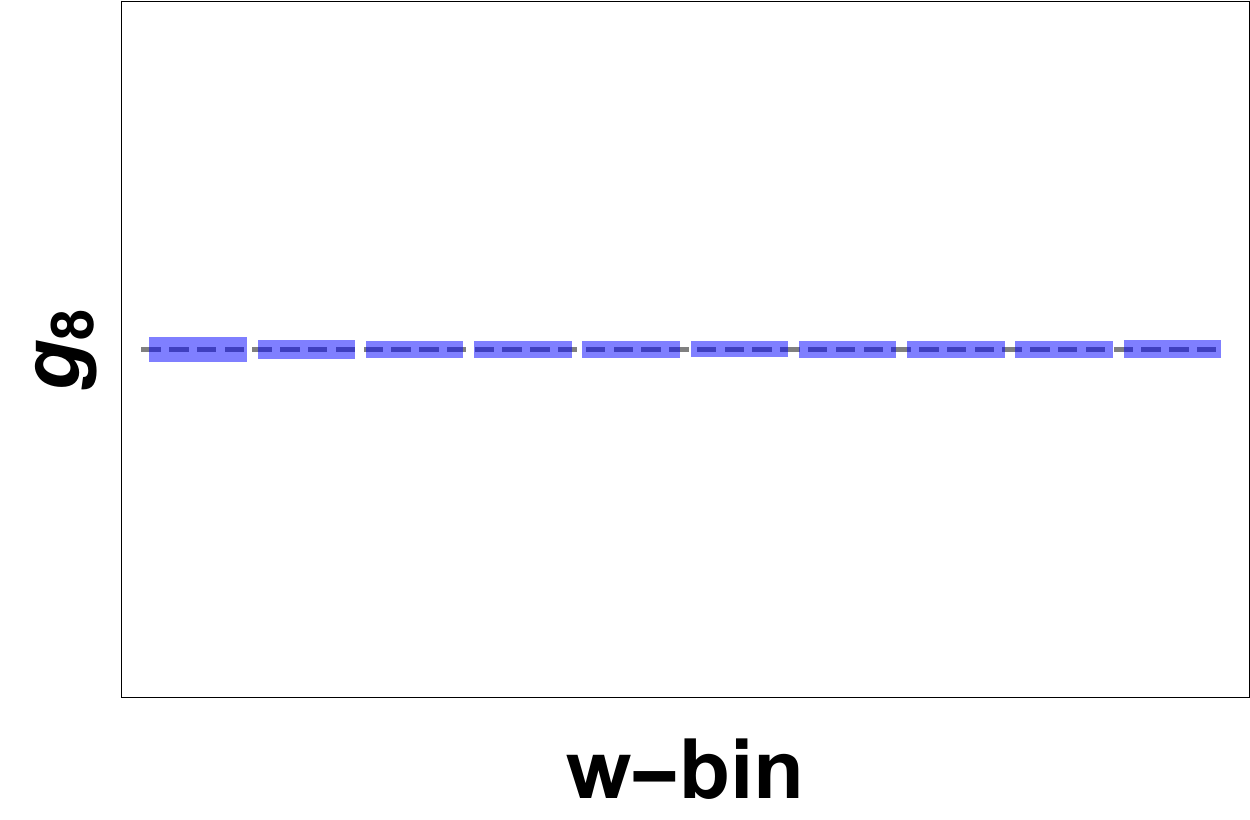}
\includegraphics[width=\figwid cm]{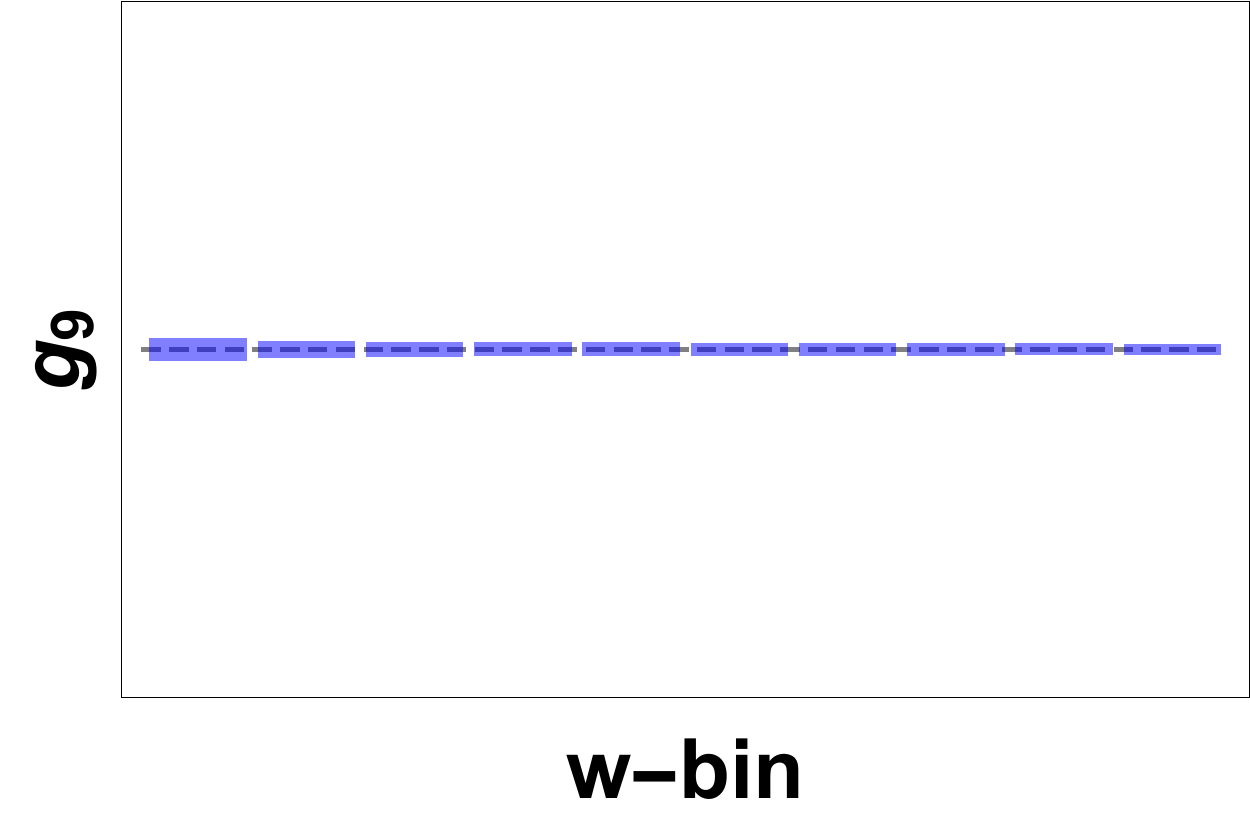}
\caption{Distribution of $\gf{i}$ in ten w-bins.}
\label{fig:gfunc_CNL}
\end{figure}

For illustration, the pseudodata with CLN parameters   
leads to the number of events for $w$-bin as
\begin{small}
\begin{eqnarray}
&N_{\rm event} =(5306, 8934, 10525, 11241, 11392, 11132, & \nn \\
&10555, 9726, 8693, 7497)\,.&  \label{eq:29}
\end{eqnarray}
\end{small}
Similarly, using the pseudodata with BGL parameters
we find
\begin{small}
\begin{eqnarray}
&N_{\rm event}=(5239, 8868, 10500, 11264, 11455, 11217, & \nn \\
&10638, 9776, 8676, 7368)\,.&
\end{eqnarray}
\end{small}
As an example, we show the $\gf{i}$ distribution with CLN pseudodata in Fig.~(\ref{fig:gfunc_CNL}).
\section{\label{sec:3} Sensitivity to the $C_{V_R}$}
In this section, we will show that using the two sets of pseudodata discussed in the previous section, how precisely the $C_{V_R}$ parameter can be determined assuming the right-handed vector current is the only source of new physic contribution (i.e. $C_{V_L}=1$ and $C_{V_R}$ non-zero). But first, for a sanity check, we study the SM case and compare to the result in~\cite{Waheed:2018djm}. Let us start with the CLN pseudodata. We use the lattice input  $h_{A_1}(1)=0.906\pm 0.013$ as done in~\cite{Waheed:2018djm} by including the following  $\chi^2_{\rm lattice} $ term
\begin{small}
\begin{equation}
\chi^2_{\rm lattice}(v)=\left(\frac{v^{\rm lattice}-v}{\sigma_{v}^{\rm lattice}}\right)^2 \,,\label{eq:38}
\end{equation}
\end{small}
with $v=h_{A_1}(1)$.
Then, we find
\begin{small}
\begin{eqnarray}
\vec{v}&=(h_{A_1}(1), \rho^2_{D^*}, R_1(1), R_2(1), V_{cb}) &\nn \\
&=({0.906, 1.106, 1.229, 0.852, 0.0387})\,,  &\\ \nn
\sigma_{\vec{v}}&=({0.013, 0.019, 0.011, 0.011, 0.0006})\,, &\nn
\end{eqnarray}
\end{small}
using $\alpha=6.616\times 10^{18}$ and $\eta_{\rm EW}=1.006$.

The fitted values coincide well with our input from \cite{Waheed:2018djm} while we can not directly compare the errors with~\cite{Waheed:2018djm} as the experimental efficiency is not correctly taken into account here. Nevertheless, our error is $\sim$50\% smaller and a partial reason might be the unbinned analysis we have applied here. Next, we use the BGL data. The lattice data for $h_{A_1}(1)$ is again used to constrain the BGL parameter $a^f_0$ via a relation:
\begin{small}
\begin{equation}
h_{A_1}(1)=\frac{1}{2m_B \sqrt{r}P_f(0)\phi_f(0) }a_0^f\,,
\end{equation}
\end{small}
which leads to $a_0^f=0.0132\pm 0.0002$.
The fit result yields
\begin{small}
\begin{eqnarray}
&\vec{v}=({a}^f_{0}, {a}^f_{1}, {a}^{\mathcal F_1}_1, {a}^{\mathcal F_1}_2, {a}^g_{0}, V_{cb}) &\\
& =({0.0132,0.0169,0.0070,-0.0853,0.0242,0.0384})\,,  &\nn \\
&\sigma_{\vec{v}}=({0.0002,0.0028,0.0011,0.0199,0.0004,0.0006})\,, & \nn\\
\end{eqnarray}
\end{small}
using $\alpha=6.613\times10^{18}$ and $\eta_{\rm EW}=1.006$. Thus, the similar conclusion applies for the BGL case as well.

Now, let us move to our main topic, the sensitivity study of the right-handed current, $C_{V_R}$. By simply adding $C_{V_R}\neq 0$, we immediately encounter two fundamental problems; i) the fit does not converge as $C_{V_R}$ is not independent of the vector form factor, which means without knowing the SM value of the vector form factor, we can not determine $C_{V_R}$, ii) $C_{V_R}$ and $V_{cb}$ are also depending as the changes in both parameters directly impact on the branching ratio of $B\to D^*\ell \nu_\ell$. Thus, to obtain the allowed range of $C_{V_R}$, we need a precise information of the  SM value of $V_{cb}$. However, as it is manifested in the $V_{cb}$ puzzle, there is a controversy in the experimental determination of $V_{cb}$ from the semi-leptonic $b\to c\ell\nu_\ell$ transitions.

Fortunately, the first problem will be soon resolved as the lattice QCD result on the vector form factor will be available~\cite{Bazavov:2021bax}. It is important to emphasise that the right-handed current contribution can not be determined from experimental data without this lattice QCD result. For the second problem, we may also use the SM value of $V_{cb}$ obtained indirectly from the unitarity relation with other measurements. However, ignoring the $V_{cb}$ determination from the semi-leptonic decays enlarges the error on  $V_{cb}$ and the correlation between  $V_{cb}$ and $C_{V_R}$ is so strong that the obtained fit result becomes unstable, especially when we have many numbers of parameters to fit. On the other hand, we found that we can circumvent this problem entirely and determine $C_{V_R}$ at a high precision when we use only the angular distribution, i.e. $\chi^2_{\rm angle}(\vec{v})$, in which an overall factor such as $V_{cb}$ is canceled out in the normalised $\gf{i}$ functions. That is, we ignore the  $\chi^2_{w-{\rm bin}}(\vec{v})$ term, which is useful solely to determine the overall factor and the $w$ dependence of the form factors. Thus, in the following studies, we use these strategies: a) we assume that the vector form factor is known at a certain precision (we use the expected lattice QCD precision, 4\% in $R_1(1)$ and 7\% in $h_V(1)$~\cite{Kaneko:2019vkx}), b) we use only the first and the third terms of Eq.~\eqref{eq:chi2sum} and evaluate the compatibility with $V_{cb}$ after the fit.

\begin{figure}[htbp]
\begin{center}
\includegraphics[scale=0.3]{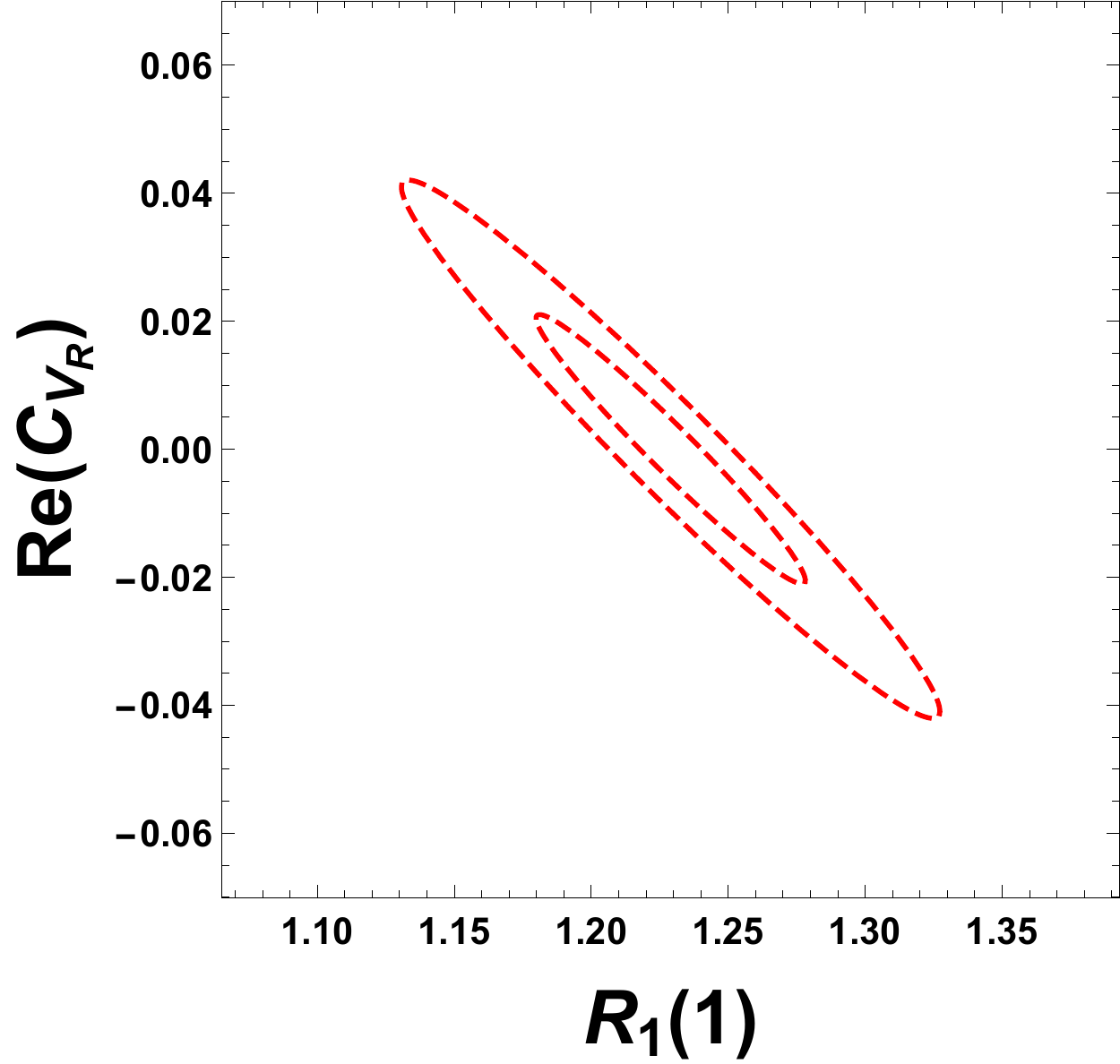}\hspace*{1cm}
\includegraphics[scale=0.3]{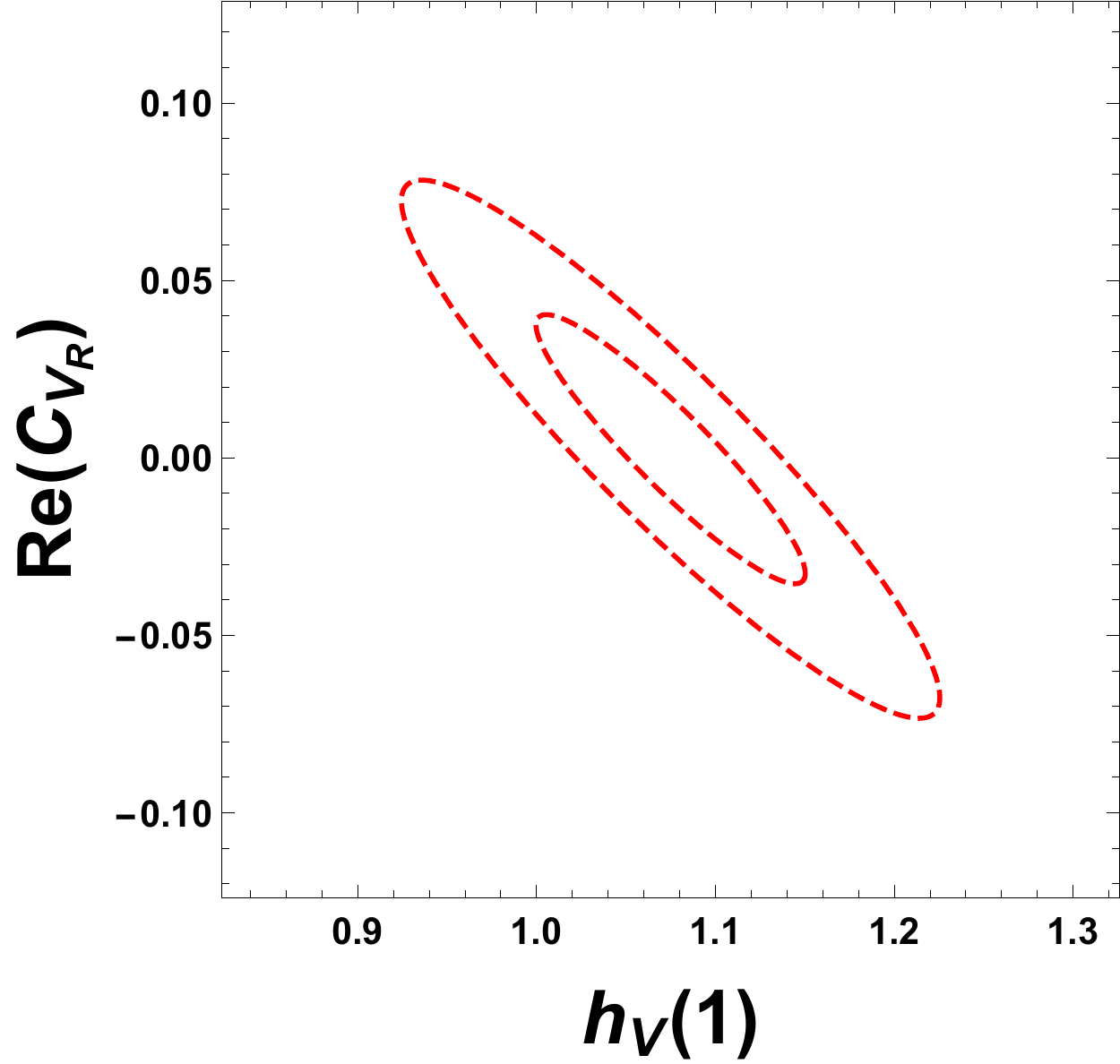}\hspace*{1cm}
\caption{$R_1(1)-C_{V_R}$ and $h_V(1)-C_{V_R}$ contours.}
\label{fig:countour}
\end{center}
\end{figure}

We first consider a scenario where $C_{V_R}$ is real , i.e. $C_{V_R}={\rm Re}(C_{V_R})$.
We start with the CLN pseudodata. The central value of the lattice input for the vector form factor, which is represented by the $R_{1}(1)$ in the CLN parametrisation,  is chosen to be our input, with 4 \% error as mentioned above, and thus, $R_{1}(1)=1.229\pm 0.049$. Our fit result yields
\begin{small}
\begin{eqnarray}
&\vec{v}=(\rho^2_{D^*}, R_1(1), R_2(1), C_{V_R}) =
({1.106,1.229,0.852,0})\,,  &\nn\\  \\
&\sigma_{\vec{v}}=({3.177,0.049,0.018,0.021}) \,,&\nn\\
&\rho_{\vec{v}}=\left(
\begin{array}{cccc}
 1. & -0.016 & -0.763 & 0.095 \\
 -0.016 & 1. & 0.006 & -0.973 \\
 -0.763 & 0.006 & 1. & -0.117 \\
 0.095 & -0.973 & -0.117 & 1. \\
\end{array}
\right)\,.
&
\end{eqnarray}
\end{small}
The most important finding here is that the $C_{V_R}$ can be determined at a 2.1\% precision. Note that the central values here is simply due to our input, where the SM is assumed, and we will know the true $C_{V_R}$ value only if an experimental data analysis is performed considering the
right-handed contribution and including the lattice QCD results on the vector form factor. In this fit, the obtained uncertainties in the form factor parameters, especially for $\rho^2_{D^*}$ and $R_2(1)$, are very large. However, the small correlation between those parameters and $C_{V_R}$ implies that ignorance of these parameters has little impact on the determination of the  $C_{V_R}$. In Fig.~\ref{fig:countour} left, we show a contour plot on the $R_1(1)-C_{V_R}$ plane. We repeat that the centre of this plot is at SM due to its initial assumption. Only when the lattice QCD value of $R_1(1)$ is obtained, we will be able to tell whether $C_{V_R}$ is deviated from the SM value, $C_{V_R}=0$, or not.
This plot shows that one day if the lattice QCD result on $R_1(1)$ becomes available and it turns out to be different from the experimental fitted value (assuming SM), non-zero $C_{V_R}$ can be hinted. We found that our result does not change significantly even if we have lattice input for $\rho^2_{D^*}, R_2(1)$, except that the errors on these parameters become smaller.

Next we study the BGL pseudodata. We again fix the vector form factor at our input value. In the BGL parametrisation, the vector form factor is related to the $a_0^g$ parameter
\begin{small}
\begin{equation}
h_{V}(1)= \frac{m_B\sqrt{r}}{P_g(0)\phi_g(0)} a_0^g\,.
\end{equation}
\end{small}
By adding the error of 7\% (larger than the ratio $R_1(1)$ which can be determined at a higher precision by lattice QCD),
we use a constraint of the $a^g_0$ parameter, $a^g_0=0.0241\pm 0.0017$. We also note that the BGL fit requires $h_{A_1}(1)$ lattice QCD input as well. The fit result yields
\begin{small}
\begin{eqnarray}
&\vec{v}=({a}^f_{0}, {a}^f_{1}, {a}^{\mathcal F_1}_1, {a}^{\mathcal F_1}_2, {a}^g_{0}, C_{V_R}) \nn \\
&=(0.0132,0.0169,0.0070,-0.0852,0.0241,0.0024)\,,  &  \\
&\sigma_{\vec{v}}=({0.0002,0.0109,0.0026,0.0352,0.0017,0.0379}
)\,, &\nn \\
&\rho_{\vec{v}}=\left(
\begin{array}{cccccc}
 1. & 0.022 & 0.039 & -0.035 & 0.000 & 0.189 \\
 0.022 & 1. & 0.860 & -0.351 & 0.000 & 0.316 \\
 0.039 & 0.860 & 1. & -0.762 & 0.000 & 0.283 \\
 -0.035 & -0.351 & -0.762 & 1. & 0.000 & -0.119 \\
 0.000 & 0.000 & 0.000 & 0.000 & 1. & -0.923 \\
 0.189 & 0.316 & 0.283 & -0.119 & -0.923 & 1. \\
\end{array}
\right)\,.
&\nn\\
\end{eqnarray}
\end{small}
We find again, despite the fact that the hadronic parameters except for $a^f_0$ \& $a^g_0$ have large uncertainties, the right-handed contribution, $C_{V_R}$, is constrained at a high precision, $\sim$4 \% level. In Fig.~\ref{fig:countour} right, we show a contour plot on the $h_V(1)-C_{V_R}$  plane, which again suggests the importance of the lattice calculation of the vector form factor in the determination of $C_{V_R}$.

Now, let us discuss the impact of the right-handed contributions to the $V_{cb}$ puzzle. As mentioned earlier, the $V_{cb}$ determination is in a contradictory situation. It is determined by the inclusive method at a $\sim$2 \% precision and by the exclusive method at a $\sim$1 \% precision, while their central values are deviated by $\sim$7 \% . The most interesting question is whether the right-handed contribution fill this gap.  The reference~\cite{Crivellin:2014zpa} pointed out that to match the exclusive $B\to D^*\ell\nu_\ell$ to the inclusive one requires $C_{V_R}\simeq -5$ \% while the exclusive $B\to D\ell\nu_\ell$ requires $C_{V_R}\simeq +5$ \%. On the other hand, if we consider only these two exclusive processes, they allow  $C_{V_R}$ to be $\sim 5$ \%. Thus, ~\cite{Crivellin:2014zpa} concluded that it is difficult to explain the $V_{cb}$ puzzle by the right-handed contributions. However, if some problem is found in one of these three measurements, or the lattice QCD input utilised to obtain $V_{cb}$, the situation could be reversed.
Our proposed method using only the angular distribution allows to pin down the $C_{V_R}$ parameter at a few \% level without intervention of the controversial $V_{cb}$ determinations, including its sign as shown in Fig.~\ref{fig:countour}. Thus, this method will provide an important step towards revealing the nature of the right-handed contribution.

Now, let us investigate the role of the $w$-dependent FBA. This observable is particularly interesting to measure: it requires only one angle measurement and many experimental errors can cancel out.  As mentioned earlier, FBA is proportional to $\gf{6s}$ and for curiosity, we investigate  what constraint on $C_{V_R}$ we would obtain from this single angular observable. The fit method is the same as before: we use the vector form factor as input. The result for the CLN case yields:
\begin{eqnarray}
&\vec{v}=(\rho^2_{D^*}, R_1(1), R_2(1), C_{V_R}) = &\nn \\
&(1.106, 1.229, 0.852, 0.000)\,,  &\\
&\sigma_{\vec{v}}=({2.200, 0.049, 0.031, 0.022}) \,,&\\
&\rho_{\vec{v}}=\left(
\begin{array}{cccc}
 1. & 0.008 & -0.873 & 0.262 \\
 0.008 & 1. & -0.040 & -0.931 \\
 -0.873 & -0.040 & 1. & -0.296 \\
 0.262 & -0.931 & -0.296 & 1. \\
\end{array}
\right)\,.&
\end{eqnarray}
It is quite intriguing that the $C_{V_R}$ can be determined at a 2.2 \% precision, which is almost as good as the case where we use the full angular coefficients. For the BGL parametrisation, we find the situation is similar: FBA alone can constrain $C_{V_R}$ at a precision of $\sim$4 \%. Since this measurement can be made as a very simple extension of the work e.g. in ~\cite{Waheed:2018djm}, we highly suggest it be done in the near future.

Finally, we discuss the models in which the right-handed interaction contain the CP violating phase, i.e. $C_{V_R}$ is a complex number. The imaginary part of the $C_{V_R}$ can be determined thanks to the angular observables $\gf{8}$ and $\gf{9}$, which are the triple product observables that can detect the CP violation without having a source of a strong phase. These observables were not included in the previous analysis, e.g.~\cite{Waheed:2018djm} as they are always zero in the SM.  They are important observables determining the CP violation and they should be introduced in the future study. In our fit, we find that Im$(C_{V_R})$ can be determined at a 0.7 \% precision, both for CLN and BGL.
\section{\label{sec:6}Conclusions}
In this article, motivated by the various interesting problems observed recently in the semi-leptonic $b\to c \ell \nu$ transitions, we investigated an application of the unbinned angular analysis of the $B\to D^*\ell \nu_\ell$ and its impact on pinning down the new physics signals. We proposed the detailed processes to apply the unbinned angular analysis to search for the right-handed contributions, in the future experimental analysis. We introduced the 11 angular coefficients and their distributions in $w$-bin, i.e. the $\gf{i}$ functions. The $\gf{i}$ functions are normalised functions, in which the overall factor in the decay rate expression, including $V_{cb}$, is canceled. Therefore, we can determine the right-handed vector contributions by circumventing the problem of the controversial values of $ V_{cb}$.

We use pseudodata computed using the theoretical parameters, from both CLN and BGL parametrisations, obtained by the fit of the Belle data~\cite{Waheed:2018djm} assuming the SM, with 95k events and performed a sensitivity study of the parameters which represent the right-handed contribution, $C_{V_R}$, assuming it is the only/dominant source of new physics. The very important finding of this study is that the new physics parameter $C_{V_R}$ and the SM parameter coming from the vector form factor can not be separately measured. Fortunately, the latter can be obtained by the lattice QCD computation and the result is expected very soon. Its central value is not known yet while the precision that the current lattice QCD computation can achieve is known to be 4 \% for $R_1(1)$ and 7 \% for $h_V(1)$. Using these values, we found that the real part of the $C_{V_R}$ can be determined at a precision of 2-4 \%. Furthermore, the imaginary part of $C_{V_R}$ can also be determined once we include the two CP violating angular coefficients, which are neglected in the previous experimental analysis. We found the imaginary part of $C_{V_R}$ can be determined at a $\sim$1 \% precision.

An additional result is obtained from a sensitivity study of the well-known FBA observable to the real part of the $C_{V_R}$ parameter. The FBA turned out to be proportional to the angular coefficient $\gf{6s}$. We performed the same fit as above but with this single angular observable. In the CLN(BGL) parametrisation, we found that the FBA alone can determine the real part of the $C_{V_R}$ at a $\sim$2(4) \% precision, which is almost equally good as the full angular coefficient fit. Therefore future measurements of FBA will be particularly useful for constraining $C_{V_R}$.
\section*{Acknowledgments}
We would like to acknowledge F. Le Diberder for his collaboration at the early stage of the project. We thank T. Kaneko, P. Urquijo, D. Ferlewicz and E. Waheed for the stimulating discussions. Z.R. Huang and E. Kou are supported by TYL-FJPPL. C.D. L\"u and R.Y. Tang are supported by Natural Science Foundation of China under grant Nos 11521505, 12070131001 and National Key Research and Development Program of China under Contract No. 2020YFA0406400.
\begin{appendix}
\section{\label{sec:A1}Appendix: Theoretical Framework}
We define the helicity amplitudes of left and right handed currents as follows
\begin{small}
\begin{align}
      H_{\lmd}^{\lmd_{D^*}}(q^2) =& {\bar \eps}_\mu^*(\lmd) \, \langle D^*(\lmd_{D^*}) | \cbar \gamma^\mu (1 - \gamma_5) b | \bar B \rangle \,,\nn\\
      {\hat H}_{\lmd}^{\lmd_{D^*}}(q^2) =& {\bar \eps}_\mu^*(\lmd) \, \langle D^*(\lmd_{D^*}) | \cbar \gamma^\mu (1 + \gamma_5) b | \bar B \rangle\,,
\end{align}
\end{small}
where ${\bar \eps}_\mu(\lmd)$ is the polarisation vector of the virtual $W$ boson.

The hadronic matrix elements describing the $\bar B \to D^{*}$ decay can be parameterised in terms of four Lorentz invariant transition form factors $V(q^2)$, $A_0(q^2)$, $A_1(q^2)$ and $A_2(q^2)$~\cite{Richman:1995wm}:
\begin{small}
\begin{align}\label{eq:ffs}
&  \langle D^{*} (p_{D^*},\epsilon) | \bar c \gamma_\mu  (1 - \gamma_5)  b |\bar  B(p_B) \rangle = \nn \\
   &  \frac{2 i V(q^2)}{m_B + m_{D^*}} \eps_{\mu\nu\alpha\beta}  \eps^{*  \nu}  p_{D^*}^\alpha  p_B^{\beta} - 2 m_{D^*} A_0(q^2) \frac{\eps^* \cdot q}{q^2} q_\mu \nn\\ &- (m_B+m_{D^*}) A_1(q^2) \left( \eps_\mu^* - \frac{\eps^* \cdot q}{q^2} q_\mu \right) \nn \\
  & + A_2(q^2)  \frac{\eps^* \cdot q}{m_B + m_{D^*}} \left[ ( p_B + p_{D^*})_\mu - \frac{m_B^2 - m_{D^*}^2}{q^2} q_\mu  \right]\,,
  \end{align}
\end{small}
where we use $\eps_{0123}=1$.

The non-zero helicity amplitudes $H_0\equiv H^0_0$, $H_{\pm}\equiv H^\pm_\pm$, ${\hat H}_0\equiv {\hat H}^0_0$ and ${\hat H}_{\pm}\equiv {\hat H}^\pm_\pm$ of left-handed and right-handed currents satisfy the following relations using form factors in Eq.~\eqref{eq:ffs}:
\begin{small}
\begin{align}
   \label{eq:hel1}
         &H_{\pm}(q^2) = -\hat{H}_{\mp}(q^2) = \nn \\
          &(m_B+m_{D^*}) A_1(q^2) \mp { 2m_B |{\bf p}_{D^*}| \over m_B+m_{D^*} } V(q^2) \,, \\
        & H_{0}(q^2) = -\hat{H}_{0}(q^2) = \nn \\
    &    { m_B+m_{D^*} \over 2m_{D^*}\sqrt{q^2} } \left[ (m_B^2-m_{D^*}^2-q^2) A_1(q^2) \right. \nn \\
                             & \left.  - { 4m_B^2 |{\bf p}_{D^*}|^2 \over (m_B+m_{D^*})^2 } A_2(q^2) \right] \,,
\label{eq:hel2}
\end{align}
\end{small}
Where $|{\bf p}_{D^*}|={\sqrt{\left[(m_B-m_{D^{*}})^2-q^2)][(m_B+m_{D^{*}})^2-q^2\right]} \over {2m_B}}$. In the following, we use $w$ variable instead of $q^2$, with $w=\frac{m_B^2+m_{D^*}^2-q^2}{2m_B m_{D^*}} $ such that $|{\bf p}_{D^*}|=m_{D^*}\sqrt{w^2-1}$ and $w=1$ corresponds to the zero-recoil momentum.

In the CLN parametrisation, the helicity amplitudes are written as
\begin{small}
\begin{eqnarray}
H_\pm(w) &=& m_B\sqrt{r}(w+1)h_{A_1}(w) \left[1\mp   \sqrt{\frac{w-1}{w+1}}R_1(w)\right]\,,\nn \\
\\
H_0(w)&=& m_B^2\sqrt{r}(w+1)\frac{1-r}{\sqrt{q^2}}h_{A_1}(w)\times\nn\\
&&\left[1+\frac{w-1}{1-r}(1-R_2(w))\right]\,,
\end{eqnarray}
\end{small}
where we have $r=m_{D^*}/m_B$ and
\begin{small}
\begin{eqnarray}
h_{A_1}(w)&=& h_{A_1}(1)(1-8\rho^2_{D^*}z+(53\rho^2_{D^*}-15)z^2\nn\\
&&~~~~~~~~~-(231\rho^2_{D^*}-91)z^3)\,, \nn\\
R_1(w)&=&R_1(1)-0.12(w-1)+0.05(w-1)^2\,, \\
R_2(w)&=&R_2(1)+0.11(w-1)-0.06(w-1)^2\,, \nn
\end{eqnarray}
\end{small}
where $z=(\sqrt{w+1}-\sqrt{2})/(\sqrt{w+1}+\sqrt{2})$.

In the BGL parametrisation, the helicity amplitudes are written as~\cite{Grinstein:2017nlq}
\begin{small}
\begin{eqnarray}
H_\pm (w)&=& f(w)\mp m_B|{\bf p}_{D^*}|g(w)\,, \\
H_0(w)&=& \frac{\mathcal{F}_1(w)}{\sqrt{q^2}}\,,
\end{eqnarray}
\end{small}
where the form factors are the expansion in terms of the $z$ variable
\begin{small}
\begin{eqnarray}
&g(z)= \frac{1}{P_g(z)\phi_g(z)}\sum_{n=0}^N a_n^gz^n\,,
f(z)= \frac{1}{P_f(z)\phi_f(z)}\sum_{n=0}^N a_n^fz^n\,,& \nn \\
&{\mathcal{F}}_1(z)= \frac{1}{P_{\mathcal{F}_1}(z)\phi_{\mathcal{F}_1}(z)}\sum_{n=0}^N a_n^{\mathcal{F}_1}z^n \,.&
\end{eqnarray}
\end{small}
The full expressions of $P_{g, f, {\mathcal{F}_1}}$ and $\phi_{g, f, {\mathcal{F}_1}}$ can be found in \cite{Waheed:2018djm}.
The  $f$ and $\mathcal{F}_1$ are not completely independent and we have
\begin{small}
\begin{equation}
\mathcal{F}_1(0)=(m_B-m_{D^*})f(0)\,,
\end{equation}
\end{small}
which leads to a relation of their leading order coefficients
\begin{small}
\begin{equation}
a_0^{\mathcal{F}_1}=(m_B-m_{D^*})\frac{\phi_{\mathcal{F}_1}(0)}{\phi_f(0)}a_0^f\,.
\end{equation}
\end{small}
We write contributions from the left-handed current, the right-handed current and the interference terms in terms of $J_i$ parameters:
\begin{small}
\begin{eqnarray}
 \label{eq:rateJ}
&&\frac{\text{d} \Gamma( \bar B \to D^{*} (\to D \pi) \, \ell^- \, \bar \nu_\ell)}{\text{d} w \, \text{d}\cos\theta_V \, \text{d}\cos\theta_{\ell} \, \text{d}\chi} \nn\\
&=& \frac{6m_Bm_{D^*}^2}{8(4\pi)^4}\sqrt{w^2-1}(1-2 \, w\, r+r^2)\, G_F^2 \, \left|V_{cb}\right|^2 \,\mathcal{B}(D^{*} \to D \pi) \nn\\
&&\times\Big\{ J_{1s} \sin^2\theta_V+J_{1c}\cos^2\theta_V +(J_{2s} \sin^2\theta_V \nn\\
&&~~ +J_{2c}\cos^2\theta_V )\cos 2\theta_\ell \nn \\
&&~~ +J_3 \sin^2\theta_V\sin^2\theta_\ell\cos 2\chi \nn \\
&&~~ +J_4\sin 2\theta_V\sin 2\theta_\ell \cos\chi
+J_5 \sin 2\theta_V\sin\theta_\ell\cos\chi \nn\\
&&~~ +(J_{6s} \sin^2\theta_V+J_{6c}\cos^2\theta_V)\cos\theta_\ell \nn \\
&&~~ +J_7 \sin 2\theta_V\sin\theta_\ell \sin\chi+J_8\sin 2\theta_V \sin 2\theta_\ell\sin\chi \nn \\
&&~~ +J_9 \sin^2\theta_V\sin^2\theta_\ell \sin2\chi\Big\}\,,
\end{eqnarray}
\end{small}
where in the massless limit of leptons the $J_i (i=1\sim 9) $ can be written by the helicity amplitudes and the Wilson coefficients of left- and right-handed currents as
\begin{small}
\begin{eqnarray}
\label{eq:Jfunction}
J_{1s}&=& \frac{3}{2}(H_+^2+H_-^2)(|C_{V_L}|^2+|C_{V_R}|^2)-6H_+H_-{\rm Re}[C_{V_L}C_{V_R}^*]\,, \nn \\
J_{1c}&=& 2H_0^2(|C_{V_L}|^2+|C_{V_R}|^2-2{\rm Re}[C_{V_L}C_{V_R}^*])\,, \nn \\
J_{2s}&=& \frac{1}{2}(H_+^2+H_-^2)(|C_{V_L}|^2+|C_{V_R}|^2)-2H_+H_-{\rm Re}[C_{V_L}C_{V_R}^*]\,, \nn \\
J_{2c}&=& -2H_0^2(|C_{V_L}|^2+|C_{V_R}|^2-2{\rm Re}[C_{V_L}C_{V_R}^*]) \,,\nn \\
J_{3}&=& -2H_+H_-(|C_{V_L}|^2+|C_{V_R}|^2)+2(H_+^2+H_-^2){\rm Re}[C_{V_L}C_{V_R}^*] \,,\nn \\
J_{4}&=& (H_+H_0+H_-H_0)(|C_{V_L}|^2+|C_{V_R}|^2-2{\rm Re}[C_{V_L}C_{V_R}^*])\,, \nn \\
J_{5}&=& -2(H_+H_0-H_-H_0)(|C_{V_L}|^2-|C_{V_R}|^2) \,,\nn \\
J_{6s}&=&-2 (H_+^2-H_-^2)(|C_{V_L}|^2-|C_{V_R}|^2)\,, \nn\\
J_{6c}&=& 0 \,,\nn\\
J_{7}&=& 0 \,,\nn\\
J_{8}&=&2 (H_+H_0-H_-H_0){\rm Im}[C_{V_L}C_{V_R}^*] \,, \nn\\
J_{9}&=&-2 (H_+^2-H_-^2){\rm Im}[C_{V_L}C_{V_R}^*] \,.
\end{eqnarray}
\end{small}
\end{appendix}


\begin{thebibliography}{99}
\bibitem{Waheed:2018djm}
E.~Waheed \textit{et al.} [Belle],
Phys. Rev. D \textbf{100}, no.5, 052007 (2019)
[erratum: Phys. Rev. D \textbf{103}, no.7, 079901 (2021)]

\bibitem{Abdesselam:2017kjf}
A.~Abdesselam \textit{et al.} [Belle],
[arXiv:1702.01521 [hep-ex]].

\bibitem{BaBar:2019vpl}
J.~P.~Lees \textit{et al.} [BaBar],
Phys. Rev. Lett. \textbf{123}, no.9, 091801 (2019)

\bibitem{Bhattacharya:2020lfm}
B.~Bhattacharya, A.~Datta, S.~Kamali and D.~London,
JHEP \textbf{07}, no.07, 194 (2020)

\bibitem{Duraisamy:2013pia}
M.~Duraisamy and A.~Datta,
JHEP \textbf{09}, 059 (2013)

\bibitem{Iguro:2020keo}
S.~Iguro, M.~Takeuchi and R.~Watanabe,
Eur. Phys. J. C \textbf{81}, no.5, 406 (2021)

\bibitem{Ivanov:2017mrj}
M.~A.~Ivanov, J.~G.~K\"orner and C.~T.~Tran,
Phys. Rev. D \textbf{95}, no.3, 036021 (2017)

\bibitem{Ivanov:2015tru}
M.~A.~Ivanov, J.~G.~K\"orner and C.~T.~Tran,
Phys. Rev. D \textbf{92}, no.11, 114022 (2015)

\bibitem{Becirevic:2016hea}
D.~Becirevic, S.~Fajfer, I.~Nisandzic and A.~Tayduganov,
Nucl. Phys. B \textbf{946}, 114707 (2019)

\bibitem{Bernlochner:2017xyx}
F.~U.~Bernlochner, Z.~Ligeti, M.~Papucci and D.~J.~Robinson,
Phys. Rev. D \textbf{96}, no.9, 091503 (2017)

\bibitem{Bernlochner:2019ldg}
F.~U.~Bernlochner, Z.~Ligeti and D.~J.~Robinson,
Phys. Rev. D \textbf{100}, no.1, 013005 (2019)

\bibitem{Bigi:2017njr}
D.~Bigi, P.~Gambino and S.~Schacht,
Phys. Lett. B \textbf{769}, 441-445 (2017)

\bibitem{Gambino:2019sif}
P.~Gambino, M.~Jung and S.~Schacht,
Phys. Lett. B \textbf{795}, 386-390 (2019)

\bibitem{Grinstein:2017nlq}
B.~Grinstein and A.~Kobach,
Phys. Lett. B \textbf{771}, 359-364 (2017)

\bibitem{Iguro:2020cpg}
S.~Iguro and R.~Watanabe,
JHEP \textbf{08}, no.08, 006 (2020)

\bibitem{Jaiswal:2020wer}
S.~Jaiswal, S.~Nandi and S.~K.~Patra,
JHEP \textbf{06}, 165 (2020)

\bibitem{Jung:2018lfu}
M.~Jung and D.~M.~Straub,
JHEP \textbf{01}, 009 (2019)

\bibitem{Bordone:2019vic}
M.~Bordone, M.~Jung and D.~van Dyk,
Eur. Phys. J. C \textbf{80}, no.2, 74 (2020)

\bibitem{Bordone:2019guc}
M.~Bordone, N.~Gubernari, D.~van Dyk and M.~Jung,
Eur. Phys. J. C \textbf{80}, no.4, 347 (2020)

\bibitem{Bobeth:2021lya}
C.~Bobeth, D.~van Dyk, M.~Bordone, M.~Jung and N.~Gubernari,
[arXiv:2104.02094 [hep-ph]].

\bibitem{Ricciardi:2019zph}
G.~Ricciardi and M.~Rotondo,
J. Phys. G \textbf{47}, 113001 (2020)

\bibitem{Colangelo:2018cnj}
P.~Colangelo and F.~De Fazio,
JHEP \textbf{06}, 082 (2018)

\bibitem{Bhattacharya:2019olg}
B.~Bhattacharya, A.~Datta, S.~Kamali and D.~London,
JHEP \textbf{05}, 191 (2019)

\bibitem{Bazavov:2021bax}
A.~Bazavov \textit{et al.} [Fermilab Lattice and MILC],
[arXiv:2105.14019 [hep-lat]].
T.~Kaneko, talk given at FPCP2021,
https://indico.ihep.ac.cn/event/12805/session/40/contribution/202 \\
The two results are in a fair agreement though we need an official average to use them in our study.

\bibitem{Fajfer:2012jt}
S.~Fajfer, J.~F.~Kamenik, I.~Nisandzic and J.~Zupan,
Phys. Rev. Lett. \textbf{109}, 161801 (2012)

\bibitem{Sakaki:2013bfa}
Y.~Sakaki, M.~Tanaka, A.~Tayduganov and R.~Watanabe,
Phys. Rev. D \textbf{88}, no.9, 094012 (2013)

\bibitem{Murgui:2019czp}
C.~Murgui, A.~Pe\~nuelas, M.~Jung and A.~Pich,
JHEP \textbf{09}, 103 (2019)

\bibitem{Becirevic:2012jf}
D.~Be\v{c}irevi\'c, N.~Ko\v{s}nik and A.~Tayduganov,
Phys. Lett. B \textbf{716}, 208-213 (2012)

\bibitem{Cheung:2020sbq}
K.~Cheung, Z.~R.~Huang, H.~D.~Li, C.~D.~L\"u, Y.~N.~Mao and R.~Y.~Tang,
Nucl. Phys. B \textbf{965}, 115354 (2021)

\bibitem{Huang:2018nnq}
Z.~R.~Huang, Y.~Li, C.~D.~Lu, M.~A.~Paracha and C.~Wang,
Phys. Rev. D \textbf{98}, no.9, 095018 (2018)

\bibitem{Li:2016vvp}
X.~Q.~Li, Y.~D.~Yang and X.~Zhang,
JHEP \textbf{08}, 054 (2016)

\bibitem{Crivellin:2012ye}
A.~Crivellin, C.~Greub and A.~Kokulu,
Phys. Rev. D \textbf{86}, 054014 (2012)

\bibitem{Altmannshofer:2017poe}
W.~Altmannshofer, P.~S.~Bhupal Dev and A.~Soni,
Phys. Rev. D \textbf{96}, no.9, 095010 (2017)

\bibitem{Asadi:2018sym}
P.~Asadi, M.~R.~Buckley and D.~Shih,
Phys. Rev. D \textbf{99}, no.3, 035015 (2019)

\bibitem{Jaiswal:2017rve}
S.~Jaiswal, S.~Nandi and S.~K.~Patra,
JHEP \textbf{12}, 060 (2017)

\bibitem{Kou:2018nap}
E.~Kou \textit{et al.} [Belle-II],
PTEP \textbf{2019}, no.12, 123C01 (2019)
[erratum: PTEP \textbf{2020}, no.2, 029201 (2020)]

\bibitem{Aaij:2015oid}
R.~Aaij \textit{et al.} [LHCb],
JHEP \textbf{02}, 104 (2016)

\bibitem{Crivellin:2009sd}
A.~Crivellin,
Phys. Rev. D \textbf{81}, 031301 (2010)

\bibitem{Crivellin:2014zpa}
A.~Crivellin and S.~Pokorski,
Phys. Rev. Lett. \textbf{114}, no.1, 011802 (2015)

\bibitem{Alioli:2017ces}
S.~Alioli, V.~Cirigliano, W.~Dekens, J.~de Vries and E.~Mereghetti,
JHEP \textbf{05}, 086 (2017)

\bibitem{Kou:2013gna}
E.~Kou, C.~D.~L\"u and F.~S.~Yu,
JHEP \textbf{12}, 102 (2013)

\bibitem{Cata:2015lta}
O.~Cat\`a and M.~Jung,
Phys. Rev. D \textbf{92}, no.5, 055018 (2015)

\bibitem{Cirigliano:2009wk}
V.~Cirigliano, J.~Jenkins and M.~Gonzalez-Alonso,
Nucl. Phys. B \textbf{830}, 95-115 (2010)

\bibitem{Bernlochner:2014ova}
F.~U.~Bernlochner, Z.~Ligeti and S.~Turczyk,
Phys. Rev. D \textbf{90}, no.9, 094003 (2014)

\bibitem{Caprini:1997mu}
I.~Caprini, L.~Lellouch and M.~Neubert,
Nucl. Phys. B \textbf{530}, 153-181 (1998)

\bibitem{Boyd:1997kz}
C.~G.~Boyd, B.~Grinstein and R.~F.~Lebed,
Phys. Rev. D \textbf{56}, 6895-6911 (1997)

\bibitem{Bernlochner:2017jka}
F.~U.~Bernlochner, Z.~Ligeti, M.~Papucci and D.~J.~Robinson,
Phys. Rev. D \textbf{95}, no.11, 115008 (2017)
[erratum: Phys. Rev. D \textbf{97}, no.5, 059902 (2018)]

\bibitem{Belle:2000cnh}
A.~Abashian \textit{et al.} [Belle],
Nucl. Instrum. Meth. A \textbf{479}, 117-232 (2002)

\bibitem{ParticleDataGroup:2020ssz}
P.~A.~Zyla \textit{et al.} [Particle Data Group],
PTEP \textbf{2020}, no.8, 083C01 (2020)

\bibitem{Kaneko:2019vkx}
T.~Kaneko \textit{et al.} [JLQCD],
PoS \textbf{LATTICE2019}, 139 (2019)

\bibitem{Richman:1995wm}
J.~D.~Richman and P.~R.~Burchat,
Rev. Mod. Phys. \textbf{67}, 893-976 (1995)
\end{thebibliography}
\end{document}